%% file: main.tex
\documentclass[lettersize,journal]{IEEEtran}
\IEEEoverridecommandlockouts
\usepackage{cite}
\usepackage{amsmath,amssymb,amsfonts}
\usepackage{graphicx}
\usepackage{textcomp,subcaption}
\usepackage{xcolor,multicol,multirow}
\usepackage{enumerate,textcomp,array,enumitem}
\usepackage{placeins}
\usepackage{threeparttable,url}
\usepackage{dblfloatfix}
\usepackage{flushend}
\usepackage{setspace}
\usepackage{soul}
\usepackage{algorithm}
\usepackage{hyperref}
\usepackage{algpseudocode}
\usepackage[running,displaymath]{lineno}

\def\BibTeX{{\rm B\kern-.05em{\sc i\kern-.025em b}\kern-.08em
    T\kern-.1667em\lower.7ex\hbox{E}\kern-.125emX}}

\begin{document}

\title{Leveraging Multi-Task Learning for Multi-Label Power System Security Assessment}
  
\author{
M.~E.~Za'ter, \IEEEmembership{Student Member, IEEE},
A.~Sajadi, \IEEEmembership{Senior Member, IEEE}, and
B.~M.~Hodge, \IEEEmembership{Senior Member, IEEE}

\thanks{Muhy Eddin Za\'ter is with the University of Colorado Boulder, USA, Corresponding Author: muhy.zater@colorado.edu}

\thanks{A.~Sajadi and B.~M.~Hodge are with the  University of Colorado Boulder, Boulder, CO 80309 USA, and the  National
Renewable Energy Laboratory (NREL), Golden, CO 80401 USA (e-mail:
Amir.Sajadi@\{colorado.edu,nrel.gov\},  BriMathias.Hodge@colorado.edu and Bri.Mathias.Hodge@nrel.gov}
}
\maketitle

\begin{abstract}
This paper introduces a novel approach to the power system security assessment using Multi-Task Learning (MTL), and reformulating the problem as a multi-label classification task. The proposed MTL framework simultaneously assesses static, voltage, transient, and small-signal stability, improving both accuracy and interpretability with respect to the most state of the art machine learning methods. It consists of a shared encoder and multiple decoders, enabling knowledge transfer between stability tasks. Experiments on the IEEE 68-bus system demonstrate a measurable superior performance of the proposed method compared to the extant state-of-the-art approaches.
\end{abstract}
\begin{IEEEkeywords}
Machine learning, multi-tasking learning, power system stability, power system security assessment
\end{IEEEkeywords}

\input{New_sections/section_1}
\input{New_sections/section_4}
\input{New_sections/section_5}
\input{New_sections/new_section_6}
\input{New_sections/section_7}

\bibliographystyle{IEEEtran}
\bibliography{Muhy_Lib}

\end{document}

%% file: New_sections/section_1.tex
\section{Introduction}
%


The power system security assessment (PSSA) is essential power application in energy management systems \cite{sajadi2019integration} apparatus that ensures the reliability and stability of energy delivery \cite{alimi2020review}. Power system operators routinely perform security assessments to ensure the system can withstand disturbances, typically involving steady-state and dynamic simulations every 15 minutes to prepare contingency plans for critical scenarios \cite{hailu2023techniques}. However, dynamic power system simulations present significant computational challenges, involving complex differential-algebraic equations with thousands of variables and constraints, requiring substantial computational resources and often facing numerical instabilities \cite{machowski1997power}. 

In recent years, mainly due to the ongoing changing landscape in the energy mix of electricity grids around the globe, conducting real-time PSSA has become more complex to the point that many power utilities may abandon this critical function. Instead, they rely solely on static security assessment, risking blackout as a result of dynamic instabilities. In addition, systematically checking $N-k$ contingencies is computationally infeasible for large-scale systems, and the safe operating region is inherently non-linear and non-convex, further complicating comprehensive analytical assessments \cite{sarkar2022search,gunda2016analysis}. 

To address these computational challenges, various analytical  methods were developed to eliminate the need to find the explicit solution of system full-order equations, some of which date back to late 80s, such as the Lyapunov function and Transient Energy Function (TEF), along with stability certificates \cite{pai1989energy, gholami2020fast, kenyon2020stability}. In recent years, machine learning (ML) has emerged as a promising alternative, offering significant improvements in computational speed and efficiency for real-time PSSA. Early ML approaches, such as decision trees, provide interpretability but have limitations in performance and sensitivity \cite{wehenkel1993decision}. More advanced techniques, including neural networks, ensemble methods (e.g., Random Forests, Gradient Boosting Machines), and Support Vector Machines (SVM), demonstrate better performance but generally lack interpretability \cite{murzakhanov2020neural,zhukov2019ensemble,dhandhia2020multi}. More recent ML developments incorporate deep learning architectures, such as Long-Short-Term Memory (LSTM) networks and Graph Neural Networks (GNNs), as well as reinforcement learning approaches for real-time decision-making \cite{zhou2021transient,zhang2019deep}. Hybrid ML methods, integrating deep auto-encoders as feature representation layers, have also been proposed to improve scalability, robustness, and generalization \cite{zhang2021confidence,sun2018deep}.

Despite the advancements, ML-based methods still face significant challenges for practical PSSA implementation, particularly regarding interpretability, scalability, accuracy, robustness, sensitivity, and consistency across diverse operating conditions \cite{li2019integrated}. For safety-critical applications such as PSSA, reliability and sensitivity to different types of instabilities (e.g., static, voltage, transient, and small-signal) are essential requirements that must be explicitly considered during algorithm design. Additionally, developing comprehensive training datasets that include sufficient insecure operating points remains challenging due to the inherent reliability of real-world power systems.

This paper introduces a novel method addressing these challenges by reformulating PSSA as a multi-label classification task and employing Multi-task Learning (MTL). It offers seperate but simultaneous static and dynamic security assessments due to the use of multi-task learning. Unlike previous approaches \cite{wehenkel1993decision, zhang2021confidence, sun2018deep, zhou2021transient, zhang2019deep} that assessed security either as a single generic classification task or by separately addressing individual instability types, our proposed framework simultaneously evaluates multiple stability aspects. The proposed MTL leverages shared knowledge across these tasks, enhancing both overall accuracy and interpretability. By identifying specific instability types, the approach provides actionable insights to operators, facilitating more informed decision-making. The proposed architecture utilizes a shared conditional auto-encoder for representation learning, combined with multiple decoders tailored to different stability tasks, improving scalability and robustness. Experiments on the IEEE 68-bus system demonstrate superior performance compared to existing state-of-the-art ML-based PSSA methods. Details on the dataset construction and experimental methodology are presented in Section III.

This paper is an extension to our previous work \cite{za2024semi}, in which we used multi-task learning to jointly train the auto-encoder and the classifier but as a single classification task for all security criteria. Whereas in the present paper, we reformulate the problem as a multi label classification task and utilize the one-encoder multiple decoder multi task learning framework \cite{lecun2015deep}. The new approach introduced in this paper enhances the accuracy and interpratbility of the system in addition to its generalization across unseen topologies as will be demonstrated later in this paper, which is precisely where the novelty of this paper resides. 


For validation purposes, experiments were conducted on the IEEE 68 bus system on synthetic data generated via Time-Domain-Simulations (TDS) and compared against other machine learning algorithms for Power system security assessment. We have made the scripts used for data generation and the implementation of the algorithm available to public \cite{githubGitHubHodgeLabDataGenerationScriptsPSSA}.

The remainder of this paper is structured as follows: section II presents a detailed formulation of the problem, as well as presenting the proposed method in depth, explaining the multi-label classification framework and the implementation of Multi-task Learning for PSSA. Section III outlines the experimental setup, including the datasets used and the evaluation metrics employed. Finally, Section IV presents the results and discusses the proposed method's performance compared to existing techniques.

%% file: New_sections/section_4.tex
\section{Proposed Methodology}

The proposed PSSA is founded upon reformulating this problem as a multi-label classification task and applying MTL. The MTL enables simultaneous learning across multiple related tasks—in this case, static, voltage, transient, and small-signal stability—by sharing a common representation space, which enhances generalization and efficiency. We explored the MTL in a recent paper \cite{za2024semi} to jointly learn the feature extraction component and the classifier. The present paper is particularly distinct and novel as it re-frames the problem as multi-label classification, and uses multi-task learning via the one encoder-multiple decoders paradigm. We suggest that this novel approach  offers a simultaneous assessment of the different security criteria with shared layer (encoder) between the security criteria as will be detailed in this section.

The proposed approach integrates three key components: (1) generation of labeled data via physics-based simulations, (2) a shared conditional autoencoder to learn compact system representations, and (3) multiple task-specific decoders for stability classification. This section outlines the core elements of the proposed framework, while details of dataset construction and evaluation methodology are deferred to Section~\ref{sec:experimental-setup}.

\begin{figure}
    \centering
    \includegraphics[width=0.9\linewidth]{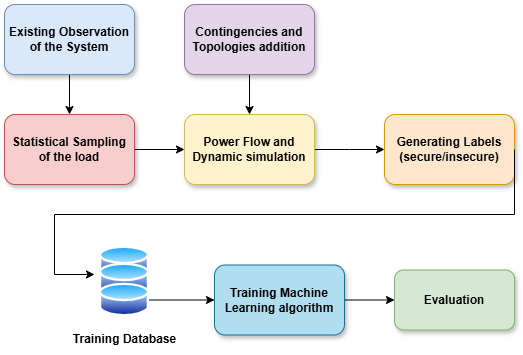}
    \caption{High-level overview of the proposed MTL-based PSSA framework.}
    \label{fig:overview}
\end{figure}

Figure~\ref{fig:overview} provides a high-level overview of the framework. We first construct a synthetic dataset by simulating multiple operating points and fault scenarios. Each operating condition is then encoded into a low-dimensional latent vector via the shared encoder. This latent representation is subsequently passed to four independent decoders, each trained to predict one type of stability outcome.

\subsection{PSSA as Multi-Label Classification}
 
The fundamental premise of MTL is that tasks are interconnected and knowledge acquired for one task can enhance performance on others, leading to more robust and efficient learning. MTL approaches have also shown significant performance enhancement for safety critical tasks that requires high sensitivity \cite{kendall2018multi}. We formulate PSSA as a multi-label classification problem, where each power system operating condition is simultaneously assessed based on four critical criteria, which is distinct from previous work of having the different security aspects as a single label \cite{za2024semi, zhang2021confidence}.

\begin{itemize}
\item \textbf{Transient Stability:} Phase angle stability following a contingency is critical in ensuring stable flow of power across a power network. For each specific contingency, the system is considered transiently insecure if the transient stability index (TSI) defined by \ref{TSI} is less than 10\%. In equation \ref{TSI}, $\triangledown\delta_{max}$ is the maximum angular separation between any two rotor angles in degrees. The TSI in Equation \ref{TSI} is based on the TSAT power swing-based algorithm \cite{liu2018accurate}.

\begin{equation}
\label{TSI}
    TSI = \frac{360-\triangledown\delta_{max}}{360+\triangledown\delta_{max}} * 100%
\end{equation}

\item \textbf{Small-signal stability:} For small-signal stability, specific performance metrics could be used as defined by the regional planning entity in which the power system of interest operates. These metrics may address damping, inertia, transient time constants, or other metrics. In this paper, we follow adopt a common industry criterion,
such as that used by Entergy—a major electric utility in
the United States—requires a minimum damping ratio of
3\% for oscillatory modes \cite{genc2010decision}. This criterion is applied to the inter-area oscillation modes of generator rotor angles, with a frequency range of 0.25-1.0 Hz and varying amplitudes.
\item  \textbf{Voltage Stability:} For voltage stability \cite{greene1999contingency}, we rely on the steady-state bifurcation curve. This criterion suggests that an increase in load condition may move the operating point closer to a bifurcation threshold. Such migration manifests itself in the form of voltage drop. Similarly, if a system is unable to adequately regulate its reactive power, voltages tend to rise. Following these principles, we elect to consider a system insecure if any bus voltage deviates from the range of 0.8 pu to 1.1 pu for more than 0.5 seconds \cite{liu2013systematic}.
\item \textbf{Static Security:} The thermal limits of the transmission line are an underlying constraint in power system operation \cite{sajadi2016great}. Although they can operate for a short period of time at their emergency rating, it is not advised to overload them for a sustained period of time. The overload index, as calculated in equation \ref{static}  \cite{sevilla2015static}, was taken into account.

\begin{equation}
\label{static}
    f_x = \sum_{i=1}^{N_l}wf_i(\frac{S_{mean,i}}{S_{max,i}})^p
\end{equation}
where $f_x$ represents the overload performance index for the operating point x, $N_l$ denotes the total number of transmission lines, and $S_{mean,i}$ and $S_{max,i}$ indicate the average and maximum apparent power flows of the $i$-th line.

\end{itemize}

This framework is intended for enhanced stability assessment, thus insecurity in any one of these aspects results in a classification of system as "insecure". This approach allows for a more holistic evaluation of system stability, capturing the inter-dependencies between different types of instabilities, and capturing the similarities within stable operating conditions.

\subsection{Architecture}

\begin{figure}[!h]
    \centering
    \includegraphics[width=0.45\textwidth]{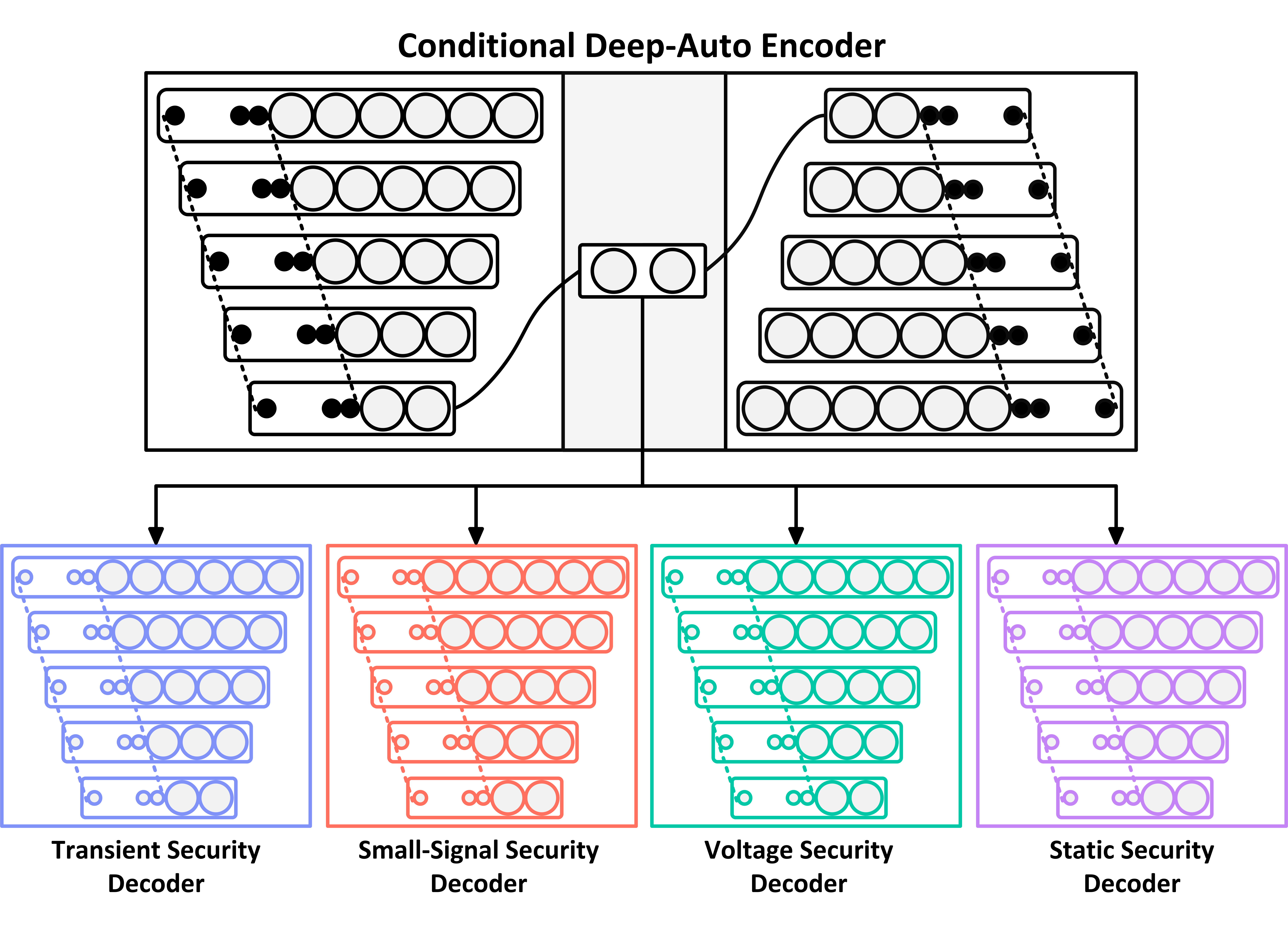}
    \caption{Encoder-decoder architecture of Multi-task Learning for PSSA}
    \label{fig:MTL}
\end{figure}

Our proposed architecture follows a one-to-many encoder-decoder paradigm, consisting of two main components\cite{lecun2015deep}. First, a shared conditional deep auto-encoder serves as the encoder, transforming raw power system features into compact vector representations for each operating condition. These representations encapsulate relevant system information and are conditioned on the applied contingencies. Second, the architecture includes multiple task-specific decoders—separate multi-layer feed-forward neural networks—each dedicated to one of the stability tasks (static, voltage, transient, or small-signal). These decoders interpret the shared latent representation to perform binary classification for their respective tasks. This architecture is illustrated in Fig.~\ref{fig:MTL}.



\subsection{Mathematical Formulation}

\subsubsection{Input Processing}
Let $x \in \mathbb{R}^d$ denote the input data representing a power system operating condition, and $c \in \mathbb{R}^k$ denote the condition vector which represents the contingencies applied. The input features include:
\begin{itemize}
\item Bus voltages: $V \in \mathbb{R}^{n_v}$
\item Line flows: $F \in \mathbb{R}^{n_f}$
\item Generator outputs: $G \in \mathbb{R}^{n_g}$
\end{itemize}
where $d = n_v + n_f + n_g$ is the total number of input features.
\subsubsection{Shared Conditional Auto-encoder}
The shared conditional auto-encoder consists of an encoder $E$ and a decoder $D$ (this decoder is separate from the task specific decoders). The encoder processes the input data and condition vector to generate a shared latent representation $h \in \mathbb{R}^m$:
\begin{equation}
h = E(x, c; \theta_E)
\end{equation}
where $\theta_E$ are the learnable parameters of the encoder. This shared representation captures common features relevant to all stability tasks, conditioned on $c$.
The encoder $E$ is composed of multiple layers as follows:
\begin{align}
h_1 &= \sigma(W_1[x;c] + b_1) \\
h_2 &= \sigma(W_2h_1 + b_2) \\
&\vdots \nonumber \\
h_L &= \sigma(W_Lh_{L-1} + b_L)
\end{align}
where $L$ is the number of layers, $W_i$ and $b_i$ are the weights and biases of the $i$-th layer, $\sigma(\cdot)$ is a non-linear activation function (e.g., ReLU \cite{lecun2015deep}), and $[x;c]$ denotes the concatenation of $x$ and $c$.
The decoder $D$ reconstructs the input from the latent representation and the condition vector:
\begin{equation}
\hat{x} = D(h, c; \theta_D)
\end{equation}
where $\theta_D$ are the learnable parameters of the decoder, and $\hat{x} \in \mathbb{R}^d$ is the reconstructed input.

The decoder $D$ is also composed of multiple layers:
\begin{align}
h'_1 &= \sigma(W'_1[h;c] + b'_1) \\
h'_2 &= \sigma(W'_2h'_1 + b'_2) \\
&\vdots \nonumber \\
\hat{x} &= \sigma(W'_M h'_{M-1} + b'_M)
\end{align}
where $M$ is the number of layers in the decoder, $W'_i$ and $b'_i$ are the weights and biases of the $i$-th layer of the decoder.
The loss function for training the conditional auto-encoder includes a reconstruction term:
\begin{equation}
\mathcal{L}{AE}(\theta_E, \theta_D) = \mathbb{E}{(x,c) \sim p_{\text{data}}}[|x - D(E(x,c;\theta_E),c;\theta_D)|_2^2]
\end{equation}
where $|\cdot|_2$ denotes the L2 norm \cite{harrington2012machine}.
This formulation allows the auto-encoder to learn a shared representation that is conditioned on additional information. The use of an auto-encoder is highly effective in enhancing both the scalability and robustness of the algorithm; it converts the raw data of the power system to lower dimensions, useful in scalability and robustness.

\subsubsection{Task-Specific Decoders}

For each task $t \in \{1, 2, 3, 4\}$ (corresponding to static, small-signal, voltage, and transient stability), a separate decoder $D_t$ is used:

\begin{equation}
    y_t = D_t(h; \theta_{D_t})
\end{equation}

where $y_t \in [0, 1]$ is the output probability for task $t$ (representing the likelihood of instability), and $\theta_{D_t}$ are the learnable parameters of the decoder for task $t$.

Each decoder $D_t$ is a multi-layer feed-forward neural network:

\begin{align}
    z_1^t &= \sigma(W_1^th + b_1^t) \\
    z_2^t &= \sigma(W_2^tz_1^t + b_2^t) \\
    &\vdots \nonumber \\
    y_t &= \sigma(W_K^tz_{K-1}^t + b_K^t)
\end{align}
where $K$ is the number of layers in the decoder, $W_i^t$ and $b_i^t$ are the weights and biases of the $i$-th layer for task $t$, and $\sigma(\cdot)$ is a non-linear activation function (with the final layer using a sigmoid activation for binary classification).

\subsubsection{Loss Function}

The overall training objective is a weighted sum of the individual task losses:

\begin{equation}
    \mathcal{L}(\theta_E, \{\theta_{D_t}\}) = \sum_{t=1}^4 \alpha_t \mathcal{L}_t(y_t, \hat{y}_t)
\end{equation}
where $\mathcal{L}_t$ is the loss function for task $t$ (typically binary cross-entropy for classification tasks), $\hat{y}_t$ is the ground truth for task $t$, and $\alpha_t$ is the weight for task $t$.

The binary cross-entropy loss for each task is defined as:

\begin{equation}
\mathcal{L}_t(y_t, \hat{y}_t) = -(\beta_t\hat{y}_t \log(y_t) + (1 - \beta_t)(1 - \hat{y}_t) \log(1 - y_t))
\end{equation}
where $\beta_t$ is the class weight for task $t$.

\subsubsection{Weight Selection}
    
We use uncertainty-based adaptive weighting \cite{kendall2018multi} where the weights are dynamically adjusted during training based on the homoscedastic uncertainty of each task:
    \begin{equation}
        \beta_t = \frac{1}{2\sigma_t^2}
    \end{equation}
    where $\sigma_t^2$ is the learned task-dependent uncertainty.
    
\subsubsection{Training Process}

The model is trained end-to-end using backpropagation to minimize the overall loss:

\begin{equation}
    \min_{\theta_E, \{\theta_{D_t}\}} \mathcal{L}(\theta_E, \{\theta_{D_t}\})
\end{equation}

This process allows the shared encoder to learn a representation that is useful for all tasks, while the task-specific decoders learn to interpret this representation for their respective stability assessments. The optimization is performed using a stochastic gradient descent variant, the Adam optimizer \cite{kingma2014adam}:

\begin{align}
    \theta_E &\leftarrow \theta_E - \eta \frac{\partial \mathcal{L}}{\partial \theta_E} \\
    \theta_{D_t} &\leftarrow \theta_{D_t} - \eta \frac{\partial \mathcal{L}}{\partial \theta_{D_t}}, \quad \forall t \in \{1,2,3,4\}
\end{align}
where $\eta$ is the learning rate, which is adapted after every epoch as in eq. \ref{adaptive}.

\begin{equation}
\label{adaptive}
    \eta_k = \eta_0 * \gamma^{k/k_s}
\end{equation}

where $\eta_k$ is the learning rate at epoch k, $\gamma$ is the decay factor, and $k_s$ is the step size.

\subsubsection{Regularization and Dropout}

To prevent overfitting, we employ L2 regularization and dropout:

\begin{equation}
    \mathcal{L}_{\text{reg}} = \mathcal{L} + \lambda \left(\|\theta_E\|_2^2 + \sum_{t=1}^4 \|\theta_{D_t}\|_2^2\right)
\end{equation}
where $\lambda$ is the regularization coefficient, and dropout is applied to the hidden layers of both the encoder and decoders during training by randomly assigning weights as zero during training.

\subsection{Advantages of the MTL Approach}

We posit that given the architecture and constituent algorithms used the proposed MTL for PSSA, it offers several key advantages:

\begin{enumerate}
    \item \textbf{Information Sharing}: By using a shared encoder, the model can capture common patterns and features relevant to multiple stability criteria, potentially leading to more efficient learning and better generalization.

    \item \textbf{Regularization}: The multi-task setup acts as a form of regularization, reducing the risk of overfitting to any single task. This is effective for PSSA as the availability of data for some security aspects, such as static security, is more straight-forward than other security tasks.

    \item \textbf{Computational Efficiency}: Instead of training separate models for each stability criterion, we train a single model that can perform all assessments simultaneously.

    \item \textbf{Inclusive Assessment}: By jointly considering multiple stability criteria, the model can potentially capture complex interactions between different types of instabilities that might be missed by separate models.

    \item \textbf{Flexibility}: The architecture allows for easy addition or removal of stability criteria by adding or removing decoder networks, making it adaptable to different power system assessment needs.
    
    \item \textbf{Interpretability}: Instead of the binary classification of secure/insecure, the multi-task learning paradigm allows for more interpretability by predicting the type of insecurity that exists in the system, providing information about potential corrective actions.
\end{enumerate}

This MTL approach to PSSA represents an enhancement over traditional single-task learning methods, offering a more comprehensive, efficient, and more accurate approach to power system security assessment.

%% file: New_sections/section_5.tex
\section{Implementation Setup}
\label{sec:experimental-setup}
\subsection{Database Construction}

The IEEE 68-bus system is used as a case study in this paper. The IEEE 68-bus power system serves as a key benchmark system that includes 16 machines distributed across five regions. This system represents a simplified equivalent of the combined New England test system (NETS) and New York power system (NYPS). The model assumes that Phasor Measurement Units (PMUs) are installed throughout the network, enabling real-time acquisition of voltage magnitudes, phase angles, and both active and reactive power at each bus. The power flow data is generated using a solver that ensures consistency with the system's physical constraints and operating conditions \cite{githubGitHubNRELSiennaPowerFlowsjl}.

To construct a comprehensive dataset for machine learning applications, operating conditions were sampled from predefined scenarios by varying the active load, ensuring a diverse representation of possible power system states. The active load power was modeled using a multivariate Gaussian distribution, with values scaled within $\pm 50\%$ of their nominal values. Reactive power scaling was performed under constant bus impedance assumptions while maintaining a power factor between 0.95 and 1. This data generation process provides realistic variations in system loading and operating conditions. The dataset captures pre-fault operating conditions, including active and reactive power for both generators and loads, power flows across transmission lines, voltage magnitudes at each bus, and phase angles. The generated power values for generation and load were denoted as $G^{\text{active}}_{\text{original}}$, $G^{\text{reactive}}_{\text{original}}$, $L^{\text{active}}_{\text{original}}$, and $L^{\text{reactive}}_{\text{original}}$, respectively, while the power flow, voltage, and phase angles were represented as $F^{\text{active}}_{\text{original}}$, $F^{\text{reactive}}_{\text{original}}$, $V_{\text{original}}$, and $\Theta_{\text{original}}$. These features collectively form the dataset $X_{\text{original}} \in \mathbb{R}^{n \times m}$, where $n$ represents the number of samples per topology and $m = 2 \times (g + l + f) + v + \theta$ represents the total number of features.

In addition to normal operating conditions, contingency scenarios were introduced to simulate the system’s response to faults. The selection of contingencies prioritized faults near generators, with fault clearing events leading to line tripping. The dataset includes variations across 18 distinct system topologies to capture a wide range of operating conditions and possible disruptions. For instance, the double line connection between bus 27 and No. 53 was identified as non-critical, as a disconnection would have minimal impact. In contrast, bus 17, which carries the system’s highest load, represents a critical point where disconnection could significantly alter power flow patterns and stability.

To evaluate the behavior of the system under these contingencies, time-domain simulations (TDS) were performed using the ANDES simulator. TDS involves numerically solving the differential-algebraic equations that govern generator dynamics, excitation systems, and network interactions over a given time horizon. This allows us to observe how system variables—such as rotor angles, voltages, and frequencies—evolve following a disturbance, and to determine whether the system recovers or becomes unstable.

The simulation results were used to generate post-fault stability labels $Y_{\text{original}} \in \mathbb{R}^{n \times 4}$. Each label vector $y_i$ represents four stability criteria—static stability, small-signal stability, voltage stability, and transient stability—where each criterion is assigned a binary value indicating stability (1) or instability (0). Mathematically, this is expressed as:

\[
y_i = [y_{i,\text{static}}, y_{i,\text{small-signal}}, y_{i,\text{voltage}}, y_{i,\text{transient}}]
\]

where:

\[
y_{i,j} = \begin{cases} 
    1, & \text{stable} \\
    0, & \text{unstable}
\end{cases}
\]

for $j \in \{\text{static, small-signal, voltage, transient}\}$.

The dataset was generated using the python packages PYPower \cite{zimmerman2010matpower} for power flow analysis, and ANDES \cite{masoom2021modelica} python wrapper for dynamic, transient analysis and small stability assessment. The codes for datasets generation are available at this \href{https://github.com/HodgeLab/DataGenerationScriptsPSSA}{link}: .

The dataset incorporates observations from multiple topologies, denoted as $T_0, T_1, \ldots, T_K$. Each topology is subjected to $C$ contingencies, expanding the dataset’s diversity and robustness. The resulting dataset is processed by normalizing and shuffling the data pairs $(X_{\text{original}}, Y_{\text{original}})$ per fault, yielding the final dataset representation where the feature matrix is $X \in \mathbb{R}^{(C \times n) \times m}$ and the corresponding labels are $Y \in \mathbb{R}^{(C \times n) \times 8}$ after one-hot encoding for each stability criterion.

Finally, the dataset is divided into training and testing subsets to enable robust model evaluation. The training and test sets are split for each topology as $X^{T_0}_{\text{train}}$, $Y^{T_0}_{\text{train}}$, $X^{T_0}_{\text{test}}$, and $Y^{T_0}_{\text{test}}$, ensuring that different topologies contribute to both sets. This structured dataset provides a rich foundation for training machine learning models to predict stability under various operating conditions and contingencies. It is worth mentioning that in section IV.D, a leave out study was conducted, where one of the topologies was removed from the training set and was included only in the testset.

\subsection{Training stage}

The training process for our multi-task learning model for PSSA is outlined in Algorithm~\ref{alg:mtl_pssa_training}. This algorithm details the steps for training the shared encoder and task-specific decoders which includes the mini batch sampling, forward and backward passes, and the loss computation.

\begin{algorithm}[!h]
\caption{Multi-Task Learning PSSA Training}
\label{alg:mtl_pssa_training}
\begin{algorithmic}[1]
\Require
\State $\mathcal{D}$: Training dataset of operating conditions and stability labels
\State $E$: Shared encoder with parameters $\theta_E$
\State $\{D_t\}_{t=1}^4$: Task-specific decoders with parameters $\{\theta_{D_t}\}_{t=1}^4$
\State $\eta$: Learning rate
\State $B$: Mini-batch size
\State $E$: Number of epochs
\State $\{\alpha_t\}_{t=1}^4$: Task weights (can be uniform, predetermined, or adaptive)
\State $\lambda$: L2 regularization coefficient

\Ensure Trained encoder $E$ and decoders $\{D_t\}_{t=1}^4$

\Function{TrainMTL-PSSA}{$\mathcal{D}, E, \{D_t\}_{t=1}^4, \eta, B, E, \{\alpha_t\}_{t=1}^4, \lambda$}
    \For{epoch $= 1$ to $E$}
        \State Shuffle $\mathcal{D}$
        \For{each mini-batch $\mathcal{B} \subset \mathcal{D}$ of size $B$}
            \State $\mathcal{L}_{\text{batch}} \gets 0$
            \For{each sample $(x, \{\hat{y}_t\}_{t=1}^4) \in \mathcal{B}$}
                \State $h \gets E(x; \theta_E)$ \Comment{Forward pass through encoder}
                \For{$t = 1$ to $4$}
                    \State $y_t \gets D_t(h; \theta_{D_t})$ \Comment{Forward pass through decoders}
                    \State $\mathcal{L}_t \gets -(\hat{y}_t \log(y_t) + (1 - \hat{y}_t) \log(1 - y_t))$
                    \State $\mathcal{L}_{\text{batch}} \gets \mathcal{L}_{\text{batch}} + \alpha_t \mathcal{L}_t$
                \EndFor
            \EndFor
            \State $\mathcal{L}_{\text{batch}} \gets \mathcal{L}_{\text{batch}} / B$ \Comment{Average loss over mini-batch}
            \State $\mathcal{L}_{\text{reg}} \gets \mathcal{L}_{\text{batch}} + \lambda (\|\theta_E\|_2^2 + \sum_{t=1}^4 \|\theta_{D_t}\|_2^2)$
            \State Compute gradients $\nabla_{\theta_E} \mathcal{L}_{\text{reg}}$ and $\nabla_{\theta_{D_t}} \mathcal{L}_{\text{reg}}$
            \State $\theta_E \gets \theta_E - \eta \nabla_{\theta_E} \mathcal{L}_{\text{reg}}$ \Comment{Update encoder parameters}
            \For{$t = 1$ to $4$}
                \State $\theta_{D_t} \gets \theta_{D_t} - \eta \nabla_{\theta_{D_t}} \mathcal{L}_{\text{reg}}$ \Comment{Update decoder parameters}
            \EndFor
        \EndFor
        \If{using adaptive weighting}
            \State Update $\{\alpha_t\}_{t=1}^4$ based on task uncertainties
        \EndIf
        \State Evaluate model on validation set and store performance metrics
        \If{early stopping criterion is met}
            \State \textbf{break}
        \EndIf
    \EndFor
    \State \Return $E, \{D_t\}_{t=1}^4$
\EndFunction
\end{algorithmic}
\end{algorithm}

\subsection{Evaluation Methodology}

To assess the performance of our multi-task learning approach for Power System Stability Assessment (PSSA), we employ a comprehensive evaluation strategy. This strategy emphasizes the critical nature of missed alarms in power systems and explores the impact of different weight selection methods in multi-task learning.

\subsubsection{Performance Metrics}

Our primary performance metric is the F2-score, which places more emphasis on recall (sensitivity) than precision. This choice reflects the greater importance of avoiding missed alarms (false negatives) in power system stability assessment, as undetected instabilities can have severe consequences.

The F2-score is defined as:

\begin{equation}
    \text{F}2 = 5 \cdot \frac{\text{precision} \cdot \text{recall}}{4 \cdot \text{precision} + \text{recall}}
\end{equation}

We compute the F2-score for two levels of assessment:

\begin{enumerate}
    \item \textbf{Overall Stability Classification}: An F2-score is calculated for the binary classification of stable vs. unstable operating conditions, considering all stability criteria collectively.
    
    \item \textbf{Individual Stability Criteria}: Separate F2-scores are computed for each of the four stability criteria (static, small-signal, voltage, and transient stability).
\end{enumerate}

\subsubsection{Comparative Analysis}

To evaluate the effectiveness of our multi-task learning approach, we compare its performance against three existing methods from the literature; decision trees \cite{wehenkel1993decision, sun2007online}, XGBoost \cite{zhukov2019ensemble} and Conditional Bayesian Deep Auto-encoder \cite{zhang2021confidence, zhang2019deep}. This comparison is performed using the same dataset and evaluation metrics to ensure a fair assessment.

%% file: New_sections/new_section_6.tex
\section{Results and Discussion}

This section presents the performance evaluation of the proposed Multi-task Learning (MTL) framework on the IEEE 68-Bus test system. A total of 18 topologies were used to generate training and testing data, with each topology subjected to multiple fault scenarios and varying operating conditions. The final testing dataset contains approximately $28,000$ samples, covering both secure and insecure cases across all four stability categories; $12,000$ insecure and $18,000$ secure samples. The data was generated via computer simulations due to the scarcity of insecure data samples in practical datasets, which hinders the ability to train a trustworthy classifier.

\subsection{Overall Performance Comparison}

\begin{figure}[!h]
    \centering
    \includegraphics[width=0.475\textwidth, height=5.25cm]{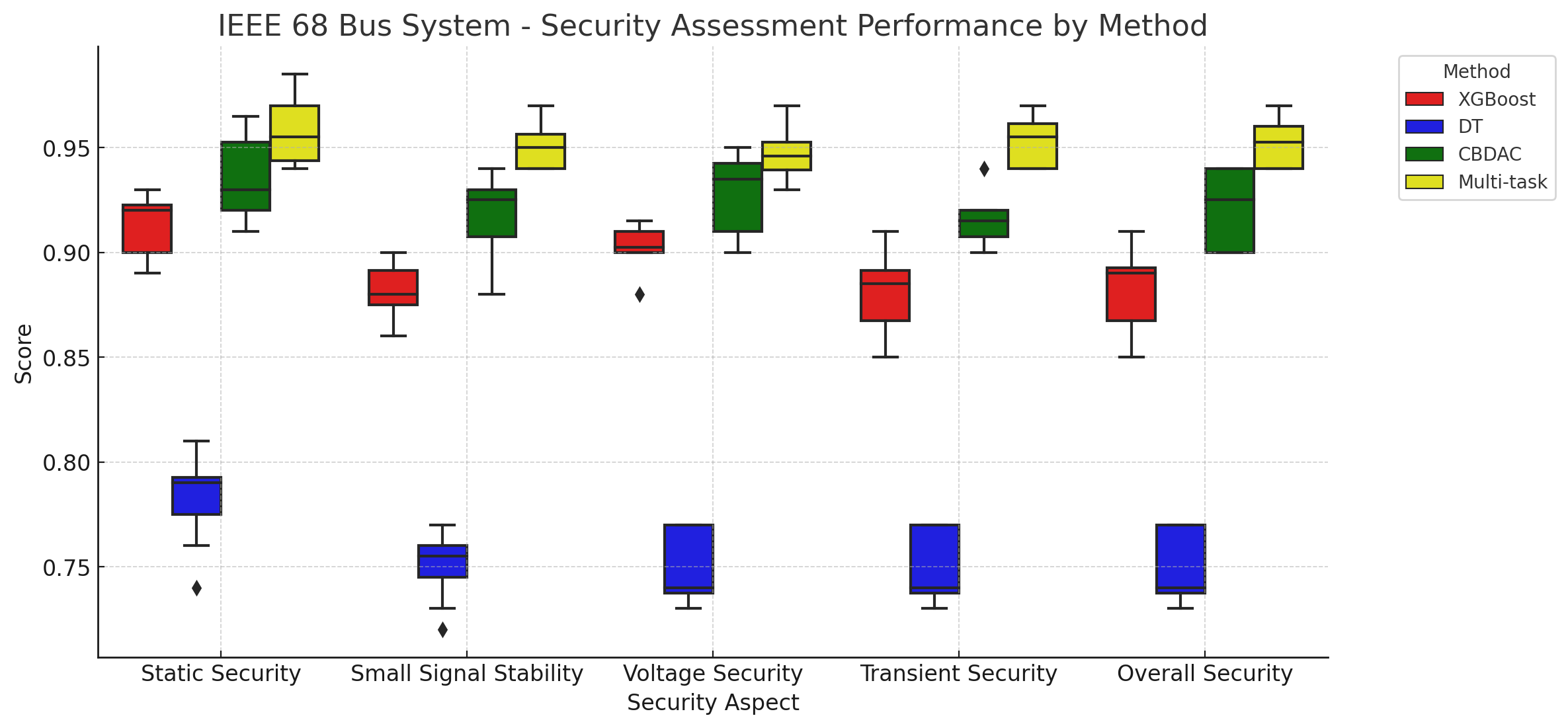}
    \caption{F2-score (weighted mean of precision and recall) for the compared methods on all security criteria on the IEEE 68 Bus system.}
    \label{fig:F2score}
\end{figure}

The comparison in Fig.~\ref{fig:F2score} shows that the proposed MTL model outperforms all other methods considered—including Decision Trees (DT) \cite{genc2010decision, wehenkel1993decision}, XGBoost \cite{zhukov2019ensemble}, and the Conditional Bayesian Deep Auto-Encoder (CBDAC) \cite{zhang2021confidence}—across all stability criteria. These methods from the literature span classical tree-based methods, ensemble learners, and deep learning architectures. MTL achieves the highest F2-scores for static, small-signal, voltage, and transient stability, as well as in the overall classification task. The superior performance is attributed to the use of a shared encoder and task-specific decoders, which enables the model to learn shared latent representations while still specializing for each task. This shared structure improves generalization and robustness, especially in transient and voltage stability predictions. Compared to the best-performing baseline method, CBDAC, the MTL model achieves an average gain of 2–3\% across individual tasks and a 3.5\% increase in overall F2-score, as well as lower variance across the different topologies.


\subsection{Precision and Recall Comparison}

To better understand model performance in terms of sensitivity and specificity, Table~\ref{tab:pr} compares the True Positive (TP) and True Negative (TN) rates across all models. The MTL method shows the highest TP and TN rates (97.78\% and 95.81\%, respectively), followed by CBDAC. This further confirms that MTL not only improves average scores but also reduces critical false negatives, which are particularly important in safety-critical systems.

\begin{table}[!h]
\centering
\caption{Comparison of True Positives and Negatives (All Criteria Combined)}
\begin{tabular}{lcc}
\hline
\textbf{Method}         & \textbf{True Positive (\%)} & \textbf{True Negative (\%)} \\ \hline
XGBoost                 & 90.76                        & 90.81                        \\
Decision Trees          & 75.36                        & 74.11                        \\
CBDAC                   & 94.23                        & 94.72                        \\
Multi-Task              & \textbf{97.78}               & \textbf{95.81}               \\ \hline
\end{tabular}
\label{tab:pr}
\end{table}

While this table aggregates performance across all security tasks, the trends are consistent with the individual task scores shown in Fig.~\ref{fig:F2score}. For instance, MTL’s advantage in transient security—a typically more difficult task—is visible both in higher F2 and higher TP/TN rates, where other models like XGBoost tend to underperform.

\subsection{Multi-task vs. Single-task Learning}

\begin{figure}[!h]
    \centering
    \includegraphics[width=0.475\textwidth, height=5.25cm]{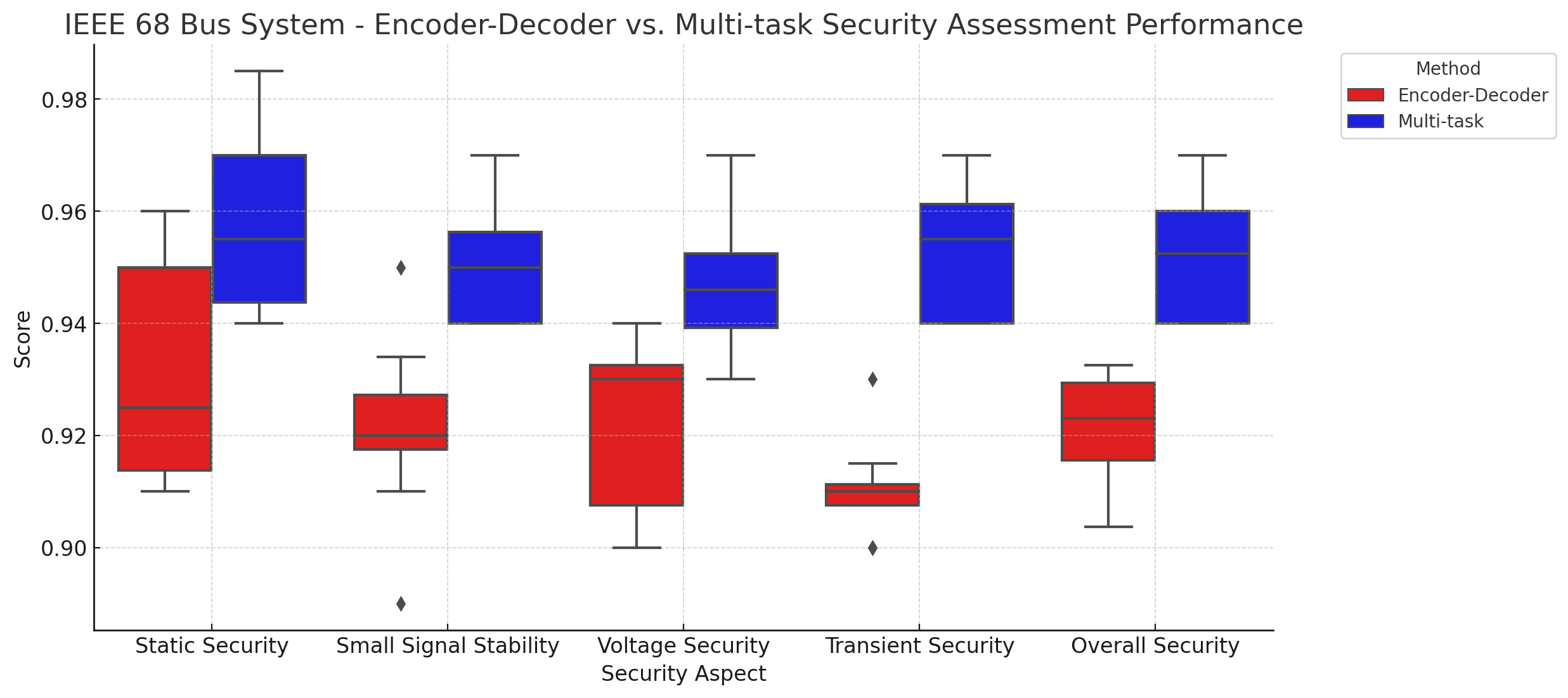}
    \caption{Performance comparison between MTL and a single-task encoder-decoder.}
    \label{fig:mtl_stl}
\end{figure}

To isolate the benefits of multi-task learning, we compare MTL to a single-task variant using the same encoder-decoder architecture but with only one decoder handling all labels jointly. The results in Fig.~\ref{fig:mtl_stl} clearly demonstrate MTL’s superiority in terms of both average score, particularly in transient and voltage security.

This supports the claim that the task-sharing mechanism in MTL leads to better generalization and improved feature extraction. Specifically, the shared encoder learns latent features beneficial across tasks (e.g., line loading, generator inertia), which results in more stable predictions. The higher variance in the single-task approach and its lower F2-scores suggest overfitting to certain classes or topologies—consistent with Fig.~\ref{fig:systematic_leave_out}, which we analyze next.

\subsection{Leave-Out Robustness Analysis}

\subsubsection{Leave-One-Category-Out Analysis}

Table~\ref{tab:leave_one_out} reports the mean overall security score when each of the four criteria is excluded during training. MTL outperforms the single-task (the classification is for a single security criteria only) method in every scenario, with the largest performance drop observed when small-signal stability is removed, confirming the interdependence of the stability aspects and the importance of considering them in an integrated manner.

\begin{table}[h]
    \centering
    \caption{Mean Overall Security Score (\%) When Excluding a Specific Category}
    \begin{tabular}{lcc}
        \hline
        \textbf{Excluded Category (\%)} & \textbf{Encoder-Decoder (\%)} & \textbf{Multi-task} \\ \hline
        Static Security            & 92.12                    & 95.19               \\
        Small Signal Stability     & 91.66                    & 94.75               \\
        Voltage Security           & 92.01                    & 94.90               \\
        Transient Security         & 92.05                    & 94.95               \\
        \hline
    \end{tabular}
    \label{tab:leave_one_out}
\end{table}

\subsubsection{Systematic Leave-One-Topology-Out Evaluation}

\begin{figure}[h]
    \centering
    \includegraphics[width=0.9\linewidth]{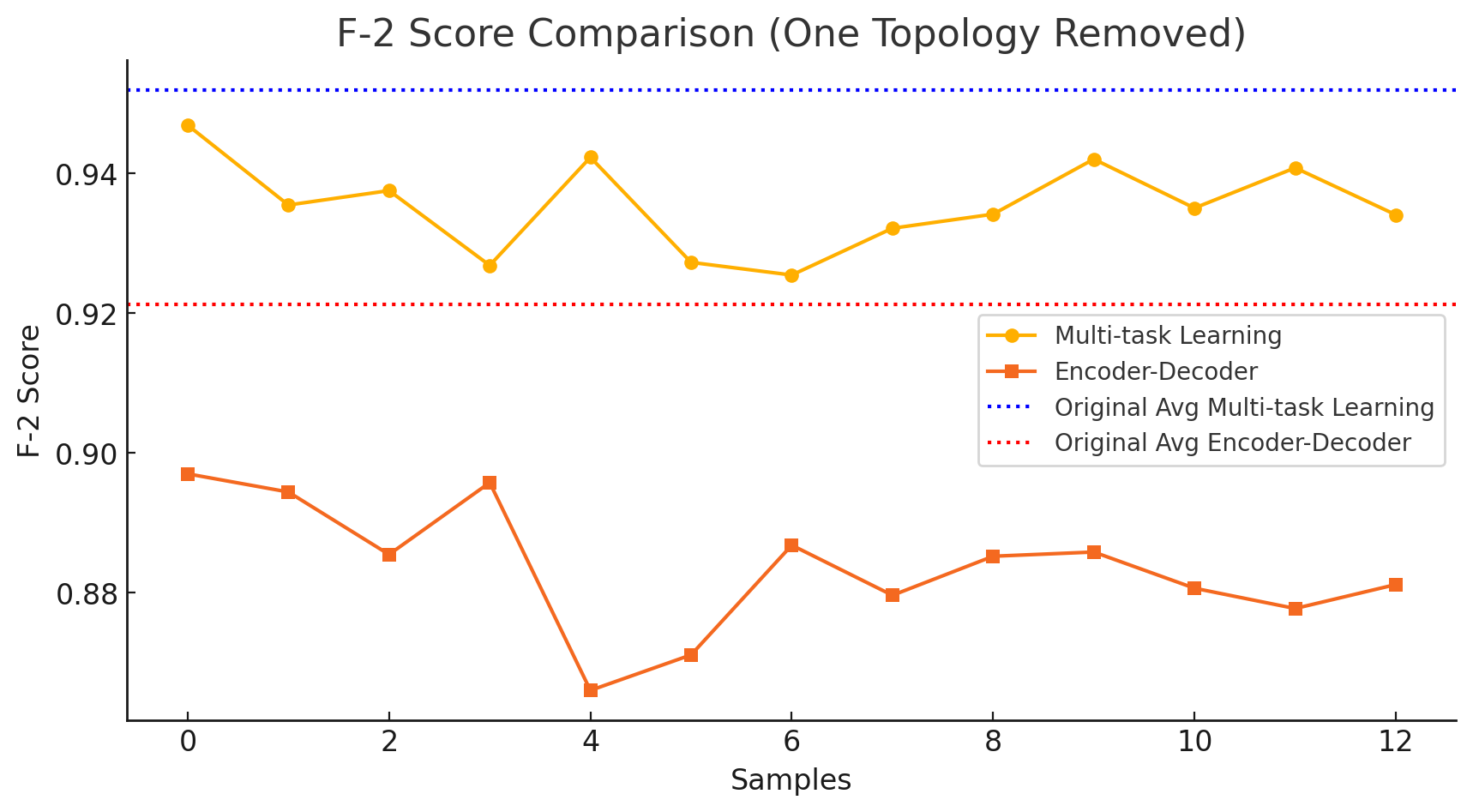}
    \caption{F2-score comparison when one topology is excluded from training.}
    \label{fig:systematic_leave_out}
\end{figure}

To evaluate robustness to unseen topologies, we removed one topology at a time from training while retaining it in testing. As shown in Fig.~\ref{fig:systematic_leave_out}, the MTL method consistently achieves higher scores and lower variability, whereas the encoder-decoder single task model model exhibits performance fluctuations. This confirms that MTL generalizes better to new grid topologies, a key requirement for practical deployment. It is worth mentioning that the reported results for the encoder-decoder single task model where aggregated from the models output on the four different security criteria.


\subsection{Discussion and Findings}

The results presented throughout this section yield the following key findings:

\begin{itemize}
    \item \textbf{Task-sharing improves generalization}: MTL's superior F2-scores and leave-out robustness (Fig.~\ref{fig:mtl_stl}, Fig.~\ref{fig:systematic_leave_out}) directly support this.
    \item \textbf{Improved feature representation}: Lower variance and higher scores in transient and voltage security suggest that MTL learns more generalizable features than the single-task model.
    \item \textbf{Overfitting in single-task models}: Fig.~\ref{fig:systematic_leave_out} shows that the single-task model over-relies on specific topologies, while MTL maintains stable performance across varying conditions.
\end{itemize}

These findings align with theoretical expectations of MTL and are empirically validated across three independent analyses: direct performance comparison, category-wise ablation, and systematic leave-out evaluation.

%% file: New_sections/section_7.tex
\section{Conclusion}

This paper presented a novel formulation of the Power System Security Assessment (PSSA) problem as a multi-label classification task, solved using a Multi-task Learning (MTL) framework. The proposed approach simultaneously evaluates static, voltage, transient, and small-signal stability, enabling a holistic and interpretable assessment of system security. This work was a natural extension of our previous work \cite{za2024semi} as this formulation and the use of multi-tasking with one to many encoder-decoder paradigm offers better interpratbility to the operators to the type of insecurity. It also offers advantages in requiring less labeled data to add a new security aspect of the power system, as the algorithm leverages the shared information between the security aspects through the use of the shared encoder, while each one has a separate decoder. Therefore the ML model performs with higher accuracy on this particular security. 

Through experiments on the IEEE 68-bus system—spanning 18 topologies and different operating conditions—we demonstrated that MTL consistently outperforms classical models (Decision Trees, XGBoost), as well as deep learning methods from the literature, such as the Conditional Bayesian Deep Auto-Encoder (CBDAC). The MTL framework achieves high F2-scores; true positive and true negative rates, and illustrates robustness under unseen topologies. These results directly validate the benefits of task-sharing, improved generalization, and robustness to variability—key requirements for deployment in real-world power systems. In addition to improved performance, the MTL architecture offers flexibility in integrating or removing security tasks, making it adaptable to different grid settings and assessment priorities. It also supports interpretability by identifying specific sources of insecurity, rather than treating system states as simply secure or insecure. Furthermore, as it can be observed from section IV.C and IV.D, when the proposed formulation was compared against the formulation of security as a single classification task (encompassing the 4 security criteria in one), it showed better overall performance on classification (figure 4) and increased generalization and robustness to removing a topology from the training as seen in figure 5.

We recognize that the proposed framework has some limitations. Its main limitation is the fact that it assumes full PMU observability and does not account for communication delays or measurement uncertainty and therefore needs to be further evaluated from this angle. We note a second limitation related to its scalability to larger systems may exist since it has not yet been tested on systems with 300 buses or more, e.g., regional power grids. In light of these limitations, future work could focus on extending this framework to partially observable settings, with integrating uncertainty quantification, and applying it to larger real-world networks with different topologies and control policies. 